\documentclass[prl,twocolumn,superscriptaddress]{revtex4}
\pdfoutput=1
\usepackage{amssymb,amsmath,amsthm,graphicx}
\usepackage{subfigure}
\usepackage{graphics, color}
\usepackage{latexsym}
\usepackage{bm}
\usepackage{epsfig}
\usepackage[pdftex]{hyperref}

\def \dd {\mathrm{d}}
\begin{document}
%---Title and Abstract ---------------------------------------------------------------
\title{The Black Ring is Unstable}
\author{Jorge E.~Santos}
\email{J.E.Santos@damtp.cam.ac.uk}
\affiliation{DAMTP, Centre for Mathematical Sciences, University of Cambridge, Wilberforce Road, Cambridge CB3 0WA, UK}
\author{Benson Way}
\email{B.Way@damtp.cam.ac.uk}
\affiliation{DAMTP, Centre for Mathematical Sciences, University of Cambridge, Wilberforce Road, Cambridge CB3 0WA, UK}
\begin{abstract}\noindent{We study non-axisymmetric linearised gravitational perturbations of the Emparan-Reall black ring using numerical methods.  We find an unstable mode whose onset lies within the ``fat" branch of the black ring and continues into the ``thin" branch.  Together with previous results using Penrose inequalities that fat black rings are unstable, this provides numerical evidence that the entire black ring family is unstable.
}
\end{abstract}
%\pacs{}
\maketitle

%---Main ---------------------------------------------------------------
%----Introduction and Summary ---------------
\noindent\emph{\bf  Introduction.}
During the golden age of general relativity, four-dimensional black holes were shown to be remarkably featureless.  The Kerr black hole is the most general asymptotically flat black hole solution to the vacuum Einstein equation \cite{PhysRevLett.11.237}. It has spherical topology and depends only on two parameters: the mass $M$ and the angular momentum $J$ \cite{Robinson:2004zz,Chrusciel:2012jk}. Moreover, for specific values of $M$ and $J$, there is only one such black hole, \emph{i.e.} the Kerr black hole is unique. Additionally, its linear stability has been shown for all modes \cite{Whiting:1988vc}. Numerical simulations suggest that Kerr black holes are also \emph{nonlinearly} stable, though a complete proof remains illusive (see however \cite{Dafermos:2010hb,Dafermos:2014cua,Dafermos:2014jwa}).  These facts have lead to the conjecture that `four-dimensional black holes have no hair'.

There is a natural higher-dimensional extension of the Kerr black hole -- the Myers-Perry black hole \cite{Myers:1986un}.  Like Kerr, it has spherical topology, depends uniquely on the mass and angular momenta, and (for sufficiently slow rotation) has substantial evidence for its stability \cite{Shibata:2010wz,Dias:2014eua}. 

Nevertheless, after the initial discovery of the black ring by Emparan and Reall \cite{Emparan:2001wn,Emparan:2006mm}, it became clear that black holes in higher dimensions are dramatically different from those in four-dimensions.  These black rings are five-dimensional solutions with topology $S^1\times S^2$ and come in two types described by their shape: ``fat" rings and ``thin" rings.  The two branches of solutions are distinct except for a unique ring that is both fat and thin. For every fat ring, there is a thin ring and a Myers-Perry black hole with the same mass and angular momentum.  Thus, uniqueness is broken among black rings, but also with Myers-Perry black holes.  In light of this, a higher-dimensional no-hair theorem seems unlikely to hold.

Yet, another surprise in higher dimensions is the existence of a new type of instability.  The so-called Gregory-Laflamme instability was first found in black strings \cite{Gregory:1993vy}, but there are analogous instabilities for rapidly rotating Myers-Perry black holes \cite{Dias:2009iu,Dias:2010eu,Shibata:2009ad,Hartnett:2013fba,Dias:2014eua}.  The general picture is that black objects with extended directions are unstable to perturbations along those extended directions. 

Since many of the solutions that violate uniqueness have extended directions, there is hope that the spirit of the no-hair theorems can be restored.  That is, there is a unique \emph{stable} solution which is a slowly rotating Myers-Perry black hole -- a dynamical no-hair conjecture.

In five dimensions, the (in)stability of the black ring is a natural setting to test this conjecture.  Indeed, they violate uniqueness, but moreover all known solutions in five dimensions are either Myers-Perry or contain a topologically $S^1\times S^2$ black object as a horizon component.  

Fat rings were conjectured to be unstable in \cite{Arcioni:2004ww,Elvang:2006dd} and confirmed in \cite{Figueras:2011he} using local Penrose inequalities. However, little is known about the stability of thin rings, except that \emph{very} thin rings ought to be unstable to Gregory-Laflamme because they resemble boosted black strings \cite{Emparan:2001wn,Emparan:2006mm,Gregory:1993vy,Hovdebo:2006jy}.  Though this argument is physically sound, it says little about the stability of thin, but not very thin, rings (which happens to coincide with where black rings violate uniqueness with themselves and Myers-Perry.)  Our goal is to demonstrate that such a window of stability does not exist by computing gravitational perturbations of the black ring.

%----Numerical Approach ---------------
\emph{\bf  Numerical Approach.}
Before embarking on a study of gravitational perturbations, let us present coordinates for the black ring that we found suitable for numerics:
	\begin{align}\label{eq:ringlineelement}
		\dd s^2&=R^2\bigg\{-(1-y^2)^2F\,\dd t^2\nonumber\\
		&\quad +\frac{{k_0}^2}{h^4}\bigg[\frac{4}{2-y^2}\frac{f_2}{g}\,\dd y^2+y^2(2-y^2)S\,(\dd\psi-\Omega h^4 W\,\dd t)^2 \nonumber\\
		&\quad+\frac{4\beta^2}{2-x^2}\frac{f_2}{f_1}\,\dd x^2+\beta^2x^2(2-x^2)(1-x^2)^2f_1\,\dd \phi^2\bigg]\bigg\}\;,
	\end{align}
where the functions $f_1$, $f_2$, $g$, $h$, $F$, $S$, and $W$ are
	\begin{align}\label{eq:functions}
	f_1&=1+\beta^2x^2(2-x^2)\;,\quad f_2=1+\alpha^2\beta^2x^2(2-x^2)\;,\nonumber\\
	g&=1+\frac{y^2(2-y^2)}{\beta^2}\;,\quad h=\sqrt{\beta^2x^2(2-x^2)+y^2(2-y^2)}\;,\nonumber\\
	F&=\frac{f_2g}{(\alpha^2 {f_1}^2-(1-y^2)^2(\alpha^2-1))g+\alpha^2(\alpha^2-1)h^4}\;,\nonumber\\
	S&=\frac{g}{F}\;,\quad W=\frac{\alpha^2(1+\alpha^2\beta^2)}{\beta^2f_2 g}F\;,
	\end{align}
the constants $\Omega$ and $k_0$ are
	\begin{equation}\label{eq:constants}
		\Omega = \frac{\beta^2\sqrt{\alpha^2-1}}{1+\beta^2}\;,\quad k_0=\frac{1+\beta^2}{\alpha\beta\sqrt{1+\alpha^2\beta^2}}\;,
	\end{equation}
and $\alpha\geq1$, $\beta>0$ are constants that parametrise this family of solutions (the constant $R$ merely sets a scale).  These parameters are related to the more familiar ones presented in \cite{Emparan:2006mm} via
	\begin{equation}
		\nu=\frac{\beta^2}{2+\beta^2}\;,\quad\lambda=\frac{\alpha^2\beta^2}{2+\alpha^2\beta^2}\;,
	\end{equation}
which satisfy $0<\nu\leq\lambda<1$.  The parameters $\lambda$ and $\nu$ are more prevalent in the literature, so we will refer to these rather than $\alpha$ and $\beta$.  The remainder of the line element in \cite{Emparan:2006mm} can be reproduced by the redefinitions
	\begin{align}
		\tilde x &= 1-2(1-x^2)^2\;,\quad \tilde y = -\frac{1-(1-\nu)(1-y^2)^2}{\nu}\;,\nonumber\\
		\tilde t&=R\,t\;,\quad \tilde\psi=\frac{\sqrt{1-\lambda}}{1-\nu}\psi\;,\quad \tilde\phi=\frac{\sqrt{1-\lambda}}{1-\nu}\phi\;,\nonumber\\
		\tilde R&=\frac{(1+\beta^2)\sqrt{2+\alpha^2\beta^2}}{\alpha\beta^2\sqrt{2+\beta^2}\sqrt{1+\alpha^2\beta^2}}R\;,
	\end{align}
where quantities with a tilde refer to those in \cite{Emparan:2006mm}.  

Our coordinates range in $x\in[0,1]$ and $y\in[0,1]$ with the axis of rotation at $y=0$, the horizon at $y=1$, the outer axis of the ring at $x=0$, and the inner axis at $x=1$.  Asymptotic infinity is at $x=y=0$, which corresponds to $h=0$.  The period of $\psi$ and $\phi$ are set to $2\pi$ and the temperature of the horizon is $1/(2\pi)$.  

Unless $\lambda=2\nu/(1+\nu^2)$, there will be a conical singularity at the inner axis.  We shall see that the singular solutions will be useful to us.  In particular, there is a static solution when $\lambda=\nu$.  We are chiefly interested in the non-singular family of solutions which we call the ``balanced" ring.  When the ring is balanced, $0<\nu\leq1/2$ are the thin rings, while $1/2\leq\nu<1$ are the fat rings.

We wish to study gravitational perturbations of \eqref{eq:ringlineelement}.  Let us write the perturbed metric as $g_{ab}={}^0g_{ab}+h_{ab}$, where ${}^0g_{ab}$ refers to the background solution \eqref{eq:ringlineelement}, and $h_{ab}$ is our metric perturbation.  Since we are working with the vacuum Einstein equation, we are free to impose the transverse-traceless gauge condition
	\begin{equation}\label{eq:transversetraceless}
		\nabla^{a}h_{ab}=0\;,\qquad h^{a}{}_a=0\;,
	\end{equation}
where, as elsewhere in this manuscript, covariant differentiation and the raising and lowering of indices are done with respect to the background ${}^0g_{ab}$.  In this gauge, the linearised Einstein equation is
	\begin{equation}\label{eq:lineareinstein}
		(\triangle_L h)_{ab}\equiv -\nabla_{c}\nabla^{c}h_{ab}-2R_{a}{}^{c}{}_{b}{}^{d}h_{cd}=0\;,
	\end{equation}
where $\triangle_L$ is the Lichnerowicz operator.  

Thin rings are extended in the $\partial_\psi$ direction, so the Gregory-Laflamme instability in the black ring would break the $\partial_\psi$ symmetry.  For simplicity, we preserve the remaining symmetry $\partial_\phi$. We therefore perform a mode decomposition $h_{ab}=e^{-i\omega t+im\psi}\tilde h_{ab}$\;, where $\tilde h_{ab}$ are functions of $x$ and $y$.

Our background has a fixed temperature $T=1/(2\pi)$, so the frequency $\omega$ is equivalent to the more general dimensionless quantity $\varpi \equiv\omega/(2\pi T)$.  Incidentally, this quantity is equivalent to $\varpi=(R_{\mathrm{out}}-R_\mathrm{in})\omega$, where $R_{\mathrm{out}}$ and $R_{\mathrm{in}}$ are the outer and inner $S^1$ equatorial horizon radii, respectively.

Since the $m=0$ modes do not break the rotation axis and $m=1$ modes were not found to be unstable in other systems with non-axisymmetric instabilities \cite{Dias:2014eua}, we will for definitiveness and simplicity set $m=2$.  

Preserving $\partial_\phi$ symmetry lets us set $h_{\mu\phi}=0$ for $\mu\neq\phi$.  This leaves 11 functions, one of which can be removed by imposing tracelessness.  After imposing tracelessness, there are six components of \eqref{eq:lineareinstein} which together with the four non-trivial components of the transverse condition in \eqref{eq:transversetraceless} form a set of 10 independent equations.  The remaining components of \eqref{eq:lineareinstein} can be derived from this set.  

Our task is to solve this set of 10, two-dimensional partial differential equations in the form of a quadratic eigenvalue problem in $\omega$.  As boundary conditions, we impose regularity on the outer axis, ingoing boundary conditions at the horizon, and outgoing boundary conditions at infinity.  On the inner axis, we demand that the conical excess/deficit does not change, which is equivalent to regularity when there is no conical singularity.

This problem is complicated by the fact that infinity is at the coordinate singularity $x=y=0$.  Because of this singularity, approaching the point $x=y=0$ from different directions will yield different values for $h_{ab}$.  To obtain a well-posed problem, our remedy is to use a different coordinate system near infinity given by
	\begin{equation}\label{eq:coordtrans}
		\rho=h\;,\quad \xi=\sqrt{1-\frac{\beta x\sqrt{2-x^2}}{h}}\;,
	\end{equation}
where the function $h$ was given in \eqref{eq:functions}. In these coordinates, spatial infinity is at the hyperslice $\rho=0$. We divide our domain into two non-overlapping coordinate patches, one in $(\rho,\xi)$ coordinates containing infinity, and another in $(x,y)$ coordinates containing the horizon and the inner axis.  Grids can then be placed on these patches using transfinite interpolation.  We must also impose additional patching conditions that require that $h_{ab}$ and its first derivatives match on patch boundaries. 

Once the problem is discretised by such a patched grid (we use pseudospectral collocation on Chebsyshev grids), it is reduced to a quadratic eigenvalue problem in linear algebra, which can be solved on a computer.  Unfortunately, because of the size of the matrices required, a direct computation of the spectrum (reduction to a linear matrix pencil followed by QZ factorisation) yields a large number of non-physical spurious modes.  We were unable to extract the physical modes from such a spectrum.

We must therefore seek an alternate route.  Rather than the entire spectrum, we wish only to obtain the unstable Gregory-Laflamme mode of the balanced ring.  By varying the parameters of the black ring, it is natural to suspect that this mode is connected to some Gregory-Laflamme mode of the \emph{static} ring, the onset of which is a zero frequency mode ($\omega=0$).  

Since we suspect that our desired mode is connected to a zero frequency mode, consider the more general problem (with the same gauge and boundary conditions)
	\begin{equation}\label{eq:negativemode}
		\triangle_Lh_{ab}=-k^2h_{ab}\;.
	\end{equation}
This problem arises in perturbations of the six-dimensional solution $(\text{black ring})\times\mathbb R$.  If there is a zero frequency mode solution to $\triangle_Lh_{ab}=0$, then there is also a zero frequency mode solution to \eqref{eq:negativemode} with $k=0$.  One can in principle find this mode by setting $\omega=0$, and solving \eqref{eq:negativemode} for $k$ while varying the parameters until $k=0$.  Indeed, this method has been used successfully in the past \cite{Dias:2009iu,Dias:2010eu}.  

For our purposes, we need not find the zero frequency mode of the static ring.  We merely observe that there are solutions of \eqref{eq:negativemode} on the static ring with $\omega=0$ and $k\neq0$ that are connected to the zero frequency mode of the static ring, and are hence also connected to our desired Gregory-Laflamme mode of the balanced ring.  

Given this, we first solve \eqref{eq:negativemode} with $\omega=0$ on the static ring for $k$. This problem is substantially simpler than solving \eqref{eq:lineareinstein} directly.  This is a \emph{linear} eigenvalue problem in $k^2$.  The static background and $\omega=0$ introduces extra symmetry and reduces the number of functions from 10 to 7.   A suitable definition of the perturbation functions also yields matrices for the eigenvalue problem that are purely real.  Furthermore, we know (as we know for black strings) that the real, positive $k^2$ modes are connected to the zero frequency mode, significantly reducing our search space for physical modes.  Attempting this problem for a particular static ring by QZ factorisation with several grid resolutions yielded a single real, positive value of $k^2$.  

Having obtained a single solution connected to the solutions we are after, we can proceed by repeated application of Newton-Raphson on several grid resolutions.  Since the Gregory-Laflamme modes of the static ring are pure imaginary, we can increase $\Gamma=-i\omega$ until we find that $k=0$.  This puts us on a solution of \eqref{eq:lineareinstein}.  From here, we can solve for $\omega$ while varying the ring parameters.  We increase the rotation until the ring is balanced and we have thus arrived on a desired solution.  We henceforth take the black ring to be balanced and present our results as we vary the parameter $\nu$.  

As a check, we have explicitly evaluated the Geroch-Held-Penrose scalars constructed in \cite{Godazgar:2012zq} and found non-zero values, confirming that this mode cannot be pure gauge. More details of our calculation and a number of numerical checks can be found in the appendix. 

\begin{figure}
\centering
\includegraphics[width=.45\textwidth]{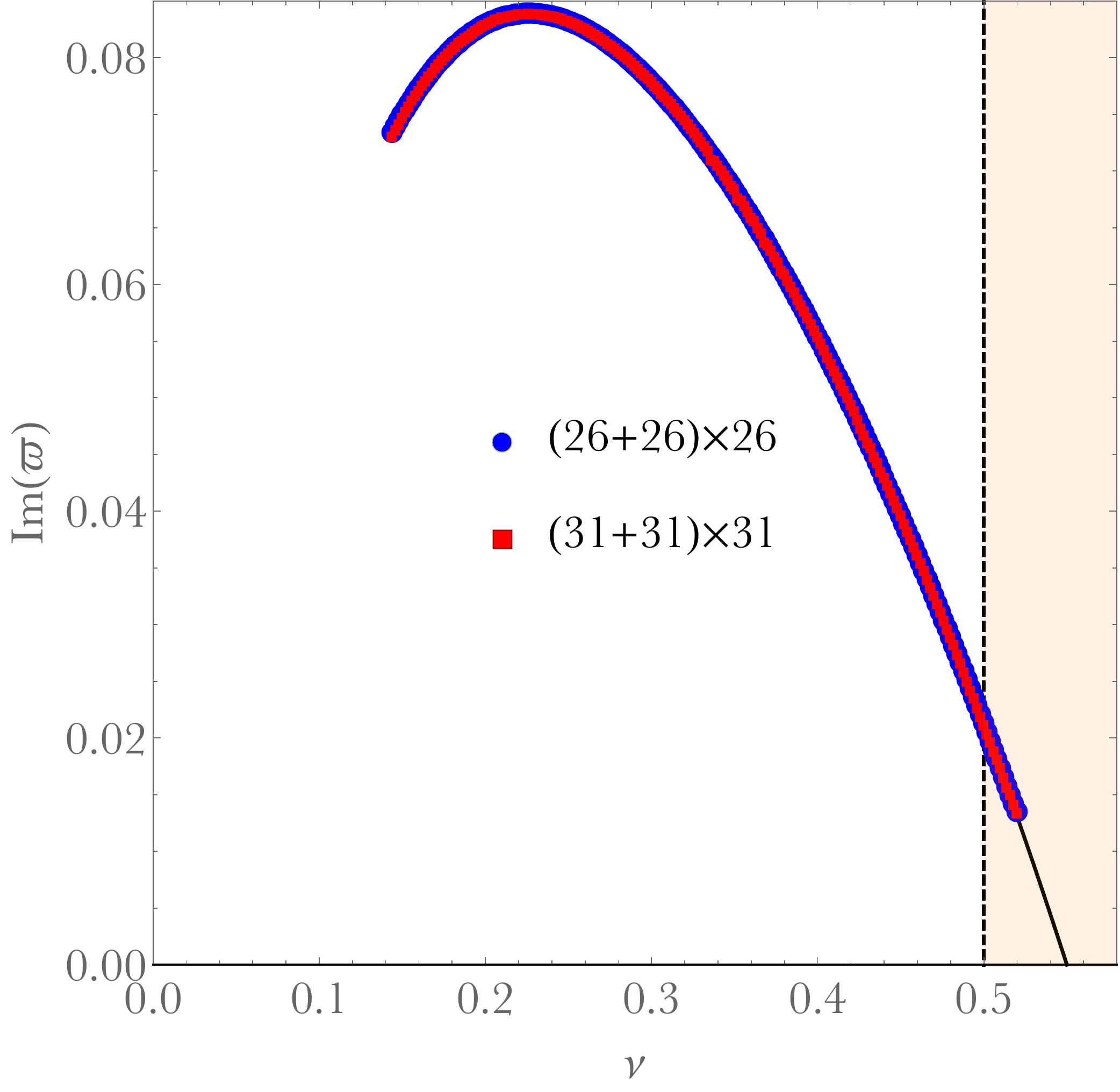}
\caption{Imaginary part of the frequency $\varpi\equiv\omega/(2\pi T)$ for non-axisymmetric perturbations with $m=2$ as a function of $\nu$ at two resolutions.  The coloured region corresponds to fat rings which have previously been shown to be unstable. The solid line is a polynomial extrapolation.}\label{fig:imomega}
\end{figure}

\begin{figure}
\centering
\includegraphics[width=.45\textwidth]{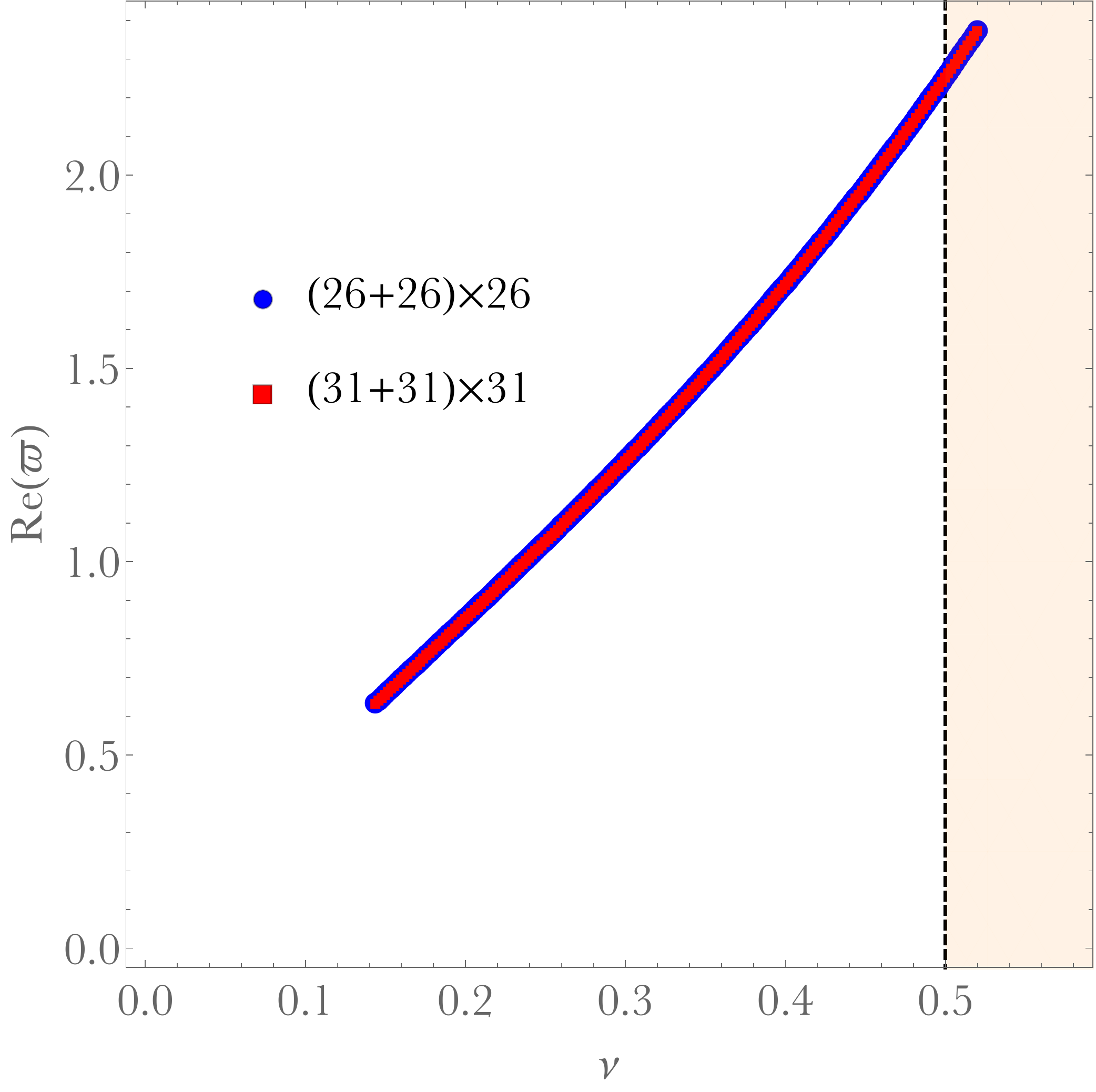}
\caption{Real part of the frequency $\varpi\equiv\omega/(2\pi T)$ as a function of $\nu$.  Same colour scheme as FIG.~\ref{fig:imomega}.}\label{fig:reomega}
\end{figure}

%----Results ---------------
\emph{\bf Results.}
FIG.~\ref{fig:imomega} and \ref{fig:reomega} present the imaginary and real parts, respectively, of the dimensionless quantity $\varpi \equiv \omega/(2\pi T)$ as a function of $\nu$.  (Recall that thin rings have $0<\nu\leq1/2$ and fat rings have $1/2\leq\nu<1$.) Dots and squares represent different resolutions, where $(N+N)\times N$ refers to $N^2$ points per patch. The agreement between resolutions is reassuring. Data points below $\nu\sim0.144$ and above $\nu\sim0.52$ were discarded due to lack of numerical precision. (Our criterion is to discard values where the two resolutions differ by more than $0.1\%$).  The vertical dashed line indicates the separation between fat and thin rings, with the coloured region corresponding to fat rings.

Our main result can be seen in Fig.~\ref{fig:imomega}, where positive $\mathrm{Im}(\varpi)$ indicates an instability. An instability seems to exist for all values of $\nu\lesssim 0.55$ (the upper bound is an extrapolation of data with $\nu>0.3$ using a second order polynomial in $\nu$).  In particular, this instability extends into a region of the fat rings with  $0.5\leq\nu\lesssim0.55$. Since all fat rings are already unstable to axisymmetric perturbations \cite{Arcioni:2004ww,Elvang:2006dd,Figueras:2011he}, this would imply that the Emparan-Reall ring is unstable for all ranges of parameters. We stress that though we do not have points below $\nu\lesssim0.144$, the Gregory-Laflamme argument should be valid for small $\nu$, so we expect the instability to persist down to $\nu=0$.

We note that the local Penrose inequalities suggest that the axisymmetric ($m=0$) instability should be marginal at $\nu=1/2$, so fat rings near $\nu=1/2$ should have a small growth rate.  Our results therefore suggest that the $m=2$ mode is dominant over the $m=0$ modes for fat rings near $\nu=1/2$.  For larger values of $\nu$, the $m=2$ modes are no longer unstable, so the $m=0$ modes should dominate.  Finding the transition point would requires data on the  $m=0$ sector of perturbations, which we leave for future work.  

A curious fact about FIG.~\ref{fig:reomega} involves the so-called superradiant bound.   Based on the results of \cite{Teukolsky:1974yv,Wald:1984rg}, the \emph{onset} of the instability must satisfy $0\leq\mathrm{Re}(\omega)\leq m\Omega$, as we have verified.  Away from the onset at $\nu \lesssim0.305$, we find instead that $\mathrm{Re}(\omega)>m\Omega$.  Though this is not in conflict with \cite{Teukolsky:1974yv,Wald:1984rg}, we do not have many examples where $\mathrm{Im}(\omega)>0$ and $\mathrm{Re}(\omega)>m\,\Omega$.  The physical significance of this property is unclear at the moment. 

\begin{figure}
\centering
\includegraphics[height=0.3\textheight]{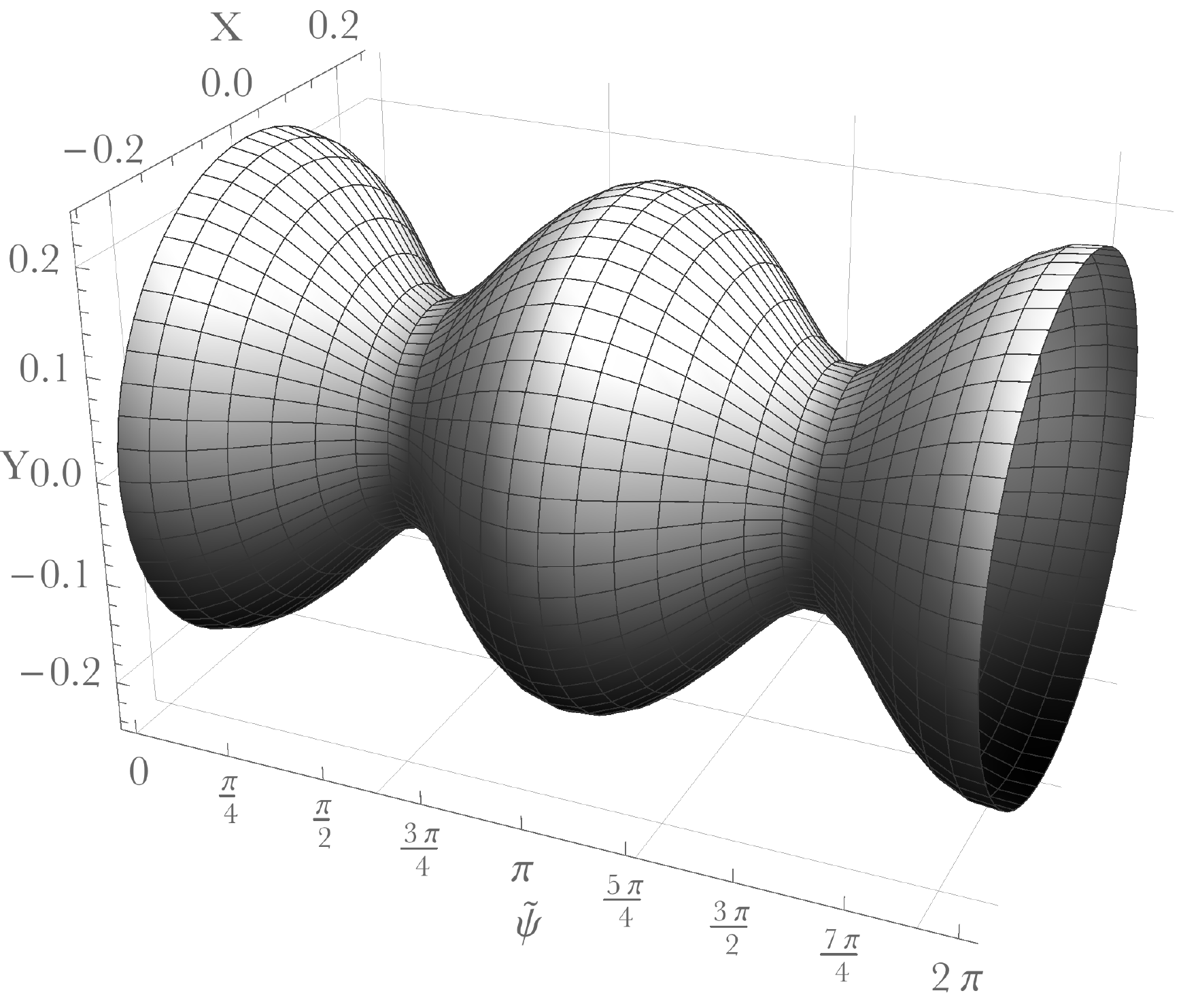}
\caption{\label{fig:embedding}Isometric embedding of constant $\psi$ slices of the $S^2$ spatial sections of the perturbed black ring horizon (see appendix for more details).  This plot corresponds to $\nu=0.2$.}
\end{figure}

To visualise the horizon evolution under this instability, we have constructed an isometric embedding for our perturbations (see appendix for more details). The results are depicted in FIG.~\ref{fig:embedding}, and are similar to those obtained in \cite{Gregory:1994bj} for the Gregory-Laflamme instability. This corroborates our claim that these non-axisymmetric perturbations are similar to that of the black string.  The general shape of the curve in Fig.~\ref{fig:imomega} is also characteristic of Gregory-Laflamme modes \cite{Gregory:1993vy,Hovdebo:2006jy}.  

%----Outlook ---------------
\emph{\bf Outlook.}
We have given substantial numerical evidence that the entire Emparan-Reall black ring is unstable.  Fat rings are unstable to axisymmetric perturbations, and thin rings are unstable to non-axisymmetric perturbations resembling the Gregory-Laflamme instability of the black string.  There is a competition between these instabilities for fat rings near $\nu=1/2$.

We focused our computation entirely on the $m=2$ modes. It would be interesting to see how the other modes behave.  A study of the $m=0$ modes would elucidate its competition with the $m=2$ modes.  There could also be instabilities corresponding to $m=1$, though these were absent in Myers-Perry \cite{Dias:2014eua}.  Perturbations with $m>2$ correspond to shorter wavelength and may have an onset for smaller $\nu$ than that of $m=2$.  One can also consider perturbations that break the $\partial_\phi$ symmetry.

While we have only studied stability of the Emparan-Reall black ring, there are many other solutions.  In five dimensions, the black ring can also rotate in the $\partial_\phi$ direction as in the double-spinning ring \cite{Pomeransky:2006bd} and the helical rings \cite{Emparan:2009vd}.  A study of rotating black strings \cite{Dias:2010eu} suggests that the instability in doubly-spinning rings would have a higher growth rate.  There are also multi-horizon solutions (\emph{e.g.} black Saturns \cite{Elvang:2007rd}, di-rings \cite{Iguchi:2007is}, and bicycling rings \cite{Elvang:2007hs}). These solutions contain black rings as horizon components and might share many of the same stability properties.  In six and higher dimensions, there are also black rings \cite{Kleihaus:2012xh,Dias:2014cia} and their associated multi-horizon solutions, but there are additionally ringoids \cite{Emparan:2009vd,Kleihaus:2014pha} and lumpy black holes \cite{Dias:2014cia,Emparan:2014pra}.  Little is known about the stability of these solutions, though many of them resemble an unstable Myers-Perry black hole, or a black ring, and hence might also share many stability properties.  The addition of matter may add a stabilising effect, particularly in supersymmetric setups \cite{Elvang:2004rt,Bena:2004de,Elvang:2004ds,Gauntlett:2004qy}.

The endpoint of these instabilities remains an important and open problem. Work is in progress \cite{pau}.  Axisymmetric instabilities are expected to lead towards Myers-Perry black holes.  For non-axisymmetric perturbations of \emph{very} thin rings, the Gregory-Laflamme instability in the (unboosted) black string suggests that black rings would develop a naked singularity and violate cosmic censorship \cite{Lehner:2010pn}.  The similarity of our results to the Gregory-Laflamme instability seems to support this idea.  It is important to note, however, that this instability in the black ring emits gravitational radiation.  Even for very thin rings, the solutions resemble \emph{boosted} black strings, and it is unclear what role this would play in the overall time evolution.

%%%%%%%%%%%%%%%%%%%%%
\newpage
\begin{center}
\emph{\bf  Acknowledgments}
\end{center}
We thank Joan Camps, Gary Horowitz and Donald Marolf for helpful discussions, \'Oscar Dias for comments on a draft of this manuscript, and Harvey Reall for comments and for being such a good sport. B.W. is supported by European Research Council grant no. ERC-2011-StG 279363-HiDGR.

\onecolumngrid
\appendix
\begin{center}
\emph{\bf  Appendix}
\end{center}

\emph{\bf{Numerical Details.}}  Here, we supply more details to our numerical approach that were absent in the main text.  Our coordinates for the background \eqref{eq:ringlineelement} are not the usual coordinates used to describe black rings such as that of \cite{Emparan:2006mm}.  It was chosen for numerical convenience (for example, many of of the boundary conditions are simpler in these coordinates).  In these coordinates, the expressions for the mass ($M$), angular momentum ($J$), temperature ($T$), horizon angular frequency ($\Omega$), and horizon area ($A$) are
	\begin{align}
		M &=\frac{3\pi R^2(1+\beta^2)^2}{8G\beta^2(1+\alpha^2\beta^2)}=\frac{3\pi R^2(1-\lambda)(1+\nu)^2}{16G(1+\lambda)\nu(1-\nu)}\rightarrow \frac{3\pi R^2}{8G\beta^2}=\frac{3\pi R^2(1-\nu)}{16G\nu}\;,\nonumber\\
		J &= \frac{\pi R^3\sqrt{\alpha^2-1}(1+\beta^2)^3}{4G\alpha^2\beta^4(1+\alpha^2\beta^2)}=\frac{\pi R^3(1-\lambda)^2(1+\nu)^3\sqrt{\lambda-\nu}}{16G\lambda(1+\lambda)(1-\nu)^2\nu^{3/2}\sqrt{1-\lambda}}\rightarrow\frac{\pi R^3(1+\beta^2)^{3/2}}{4G\beta^4(2+\beta^2)}=\frac{\pi R^3(1-\nu^2)^{3/2}}{32G\nu^2}\;,\nonumber\\
		T&=\frac{1}{2\pi R}\,\nonumber\\
		\Omega &=\frac{\beta^2\sqrt{\alpha^2-1}}{R(1+\beta^2)}=\frac{2\sqrt{\nu}\sqrt{\lambda-\nu}}{R(1+\nu)\sqrt{1-\lambda}}\rightarrow\frac{\beta^2}{R\sqrt{1+\beta^2}}=\frac{2\nu}{R\sqrt{1-\nu^2}}\;,\nonumber\\
		A&=\frac{2\pi^2R^3(1+\beta^2)^2}{\alpha^2\beta^2(1+\alpha^2\beta^2)}=\frac{\pi^2R^3(1-\lambda)^2(1+\nu)^2}{\lambda(1+\lambda)(1-\nu)^2}\rightarrow\frac{2\pi^2R^3}{2\beta^2(1+\beta^2)}=\frac{\pi^2 R^3(1-\nu)^2}{2\nu}\;,
	\end{align}
where $G$ is the gravitational constant and we have temporarily rescaled $t\rightarrow t/R$ to more clearly show how these quantities depend on the dimensionful scale $R$.  The quantities to the right of the arrows refer to the balanced ring.

Now we write down an ansatz for our perturbed metric in full:
	\begin{align}\label{eq:ansatz}
		\dd s^2&=R^2\Bigg\{-(1-y^2)^2F\left(1+\frac{\widetilde{\delta H}_1}{(1-y^2)^2}\right)\,\dd t^2\nonumber\\
		&\qquad\qquad +\frac{{k_0}^2}{h^4}\Bigg[\frac{4}{2-y^2}\frac{f_2}{g}\left(1+\frac{h^2\widetilde{\delta H}_2}{y^2(2-y^2)(1-y^2)^2}\right)\,\dd y^2+y^2(2-y^2)S\left(1+\frac{h^2\widetilde{\delta H}_4}{y^2(2-y^2)}\right)\,(\dd\psi-\Omega h^4 W\,\dd t)^2 \nonumber\\
		&\qquad\qquad\qquad\qquad+\frac{4\beta^2}{2-x^2}\frac{f_2}{f_1}(1+\widetilde{\delta H}_3)\,\dd x^2+\beta^2x^2(2-x^2)(1-x^2)^2f_1(1+\widetilde{\delta H}_5)\,\dd \phi^2\Bigg]\nonumber\\
		&\qquad\qquad+2{k_0}^2\Bigg[i\left(\frac{\widetilde{\delta H}_{10}}{y(2-y^2)(1-y^2)h^2}\,\dd y+\frac{x(1-x^2)\widetilde{\delta H}_{11}}{h^4}\,\dd x-\frac{\widetilde{\delta H}_6}{h}\,\dd t\right)(\dd\psi-\Omega h^4 W\,\dd t)\nonumber\\
		&\qquad\qquad\qquad\qquad+\frac{x(1-x^2)\widetilde{\delta H}_7}{y(2-y^2)(1-y^2)h^4}\,\dd x\,\dd y-\frac{\widetilde{\delta H}_{8}}{y(2-y^2)(1-y^2)h}\,\dd t\,\dd y-\frac{x(1-x^2)\widetilde{\delta H}_{9}}{h^3}\,\dd t\,\dd x\Bigg]\Bigg\}\;,
	\end{align}
where the $\widetilde{\delta H}$'s are unknown functions of $x$ and $y$, and the remaining functions and constants have been defined previously in \eqref{eq:functions} and \eqref{eq:constants}.  This metric is already of the form $g_{ab}={}^0g_{ab}+h_{ab}$, where ${}^0g_{ab}$ is the background solution and the $\widetilde{\delta H}$'s appear linearly in $h_{ab}$.  Now we remove the factors
	\begin{equation}\label{eq:factors}
		\widetilde{\delta H}_j = e^{-i\omega t+im\psi}e^{ik_0\omega/h}h^{3/2}(1-y^2)^{-i(\omega-m\Omega)}\left(\frac{y\sqrt{2-y^2}}{h}\right)^m\delta H_j\;,
	\end{equation}
and work with the $\delta H$'s instead.   The factor of $e^{-i\omega t+im\psi}$ is, of course, our mode decomposition of the perturbations.  The remaining factors and some of the extra factors in \eqref{eq:ansatz} correspond to the behaviour of the perturbation functions after imposing the appropriate boundary conditions.  These factors can be derived using an approach similar to that of \cite{Dias:2014eua}.  Note that some of these factors depend on $m\geq2$ and will change for $m=0$ and $m=1$.

The equations we wish to solve are
	\begin{equation}\label{eq:eomfull}
		h^a{}_a=0\;,\qquad \nabla^ah_{ab}=0\;,\qquad \triangle _Lh_{ab}=0\;.
	\end{equation}
We can use the traceless equation to eliminate $\delta H_5$ from the remaining equations.  This leaves us with 10 functions to solve for.  The linearised Einstein equation $\triangle _Lh_{ab}=0$ produces second-order equations in the $\delta H$'s of which we only keep those associated with $\delta H_j$ for $j\in\{2, 3, 7, 8, 9, 11\}$.  Together with the transverse condition $\nabla^ah_{ab}=0$, this forms our set of 10 equations that we will solve numerically.  One can verify that the remaining components of $\triangle _Lh_{ab}=0$ can be derived from this set of equations and its derivatives.  Because of the coordinate singularity at $x=y=0$, we must also obtain this same set of equations in the $(\rho,\xi)$ coordinate system defined in \eqref{eq:coordtrans}.

Now we discuss boundary conditions.  The factors we removed in \eqref{eq:factors} and our choice of ansatz \eqref{eq:ansatz} guarantee that our boundary conditions are satisfied so long as the $\delta H$'s remain finite.  Nevertheless, for numerical accuracy we impose the following boundary conditions, all of which can be derived from a series expansion of our equations.  At $x=0$, we impose
	\begin{gather}
		\delta H_1+\delta H_2+(1-y^2)^2(2\delta H_3+\delta H_4)=0\;,\nonumber\\
		\partial_x \delta H_j=0\;,\qquad j\in\{1, 2, 4, 6, 7, 8, 9, 10, 11\}\;.
	\end{gather}
At $x=1$, we impose
	\begin{gather}
		y^2(2-y^2)\delta H_1+(y^2(2-y^2)+\beta^2)\delta H_2+(1-y^2)^2(2y^2(2-y^2)\delta H_3+(y^2(2-y^2)+\beta^2)\delta H_4)=0\;,\nonumber\\
		\partial_x \delta H_j=0\;,\qquad j\in\{1, 2, 4, 6, 7, 8, 9, 10, 11\}\;.
	\end{gather}
At $y=0$, we impose
	\begin{gather}
		\delta H_2+\delta H_4=0\;,\nonumber\\
		\delta H_7-2\delta H_{11}=0\;,\nonumber\\
		\delta H_8-2\delta H_6=0\;,\nonumber\\
		\delta H_{10}+2f_2\delta H_4=0\;,\nonumber\\
		\partial_y \delta H_j=0\;,\qquad j\in\{1, 3, 4, 6, 9,11\}\;.
	\end{gather}
And at $y=1$, we impose
	\begin{gather}
		\delta H_1+f_1\delta H_2=0\;,\nonumber\\
		\delta H_2-\frac{(1+\beta^2)\sqrt{1+x^2(2-x^2)\beta^2}}{2\beta^2f_2}\delta H_{8}=0\;,\nonumber\\
		\delta H_7-2\sqrt{1+x^2(2-x^2)\beta^2}\delta H_9=0\;,\nonumber\\
		\delta H_{10}-2\sqrt{1+x^2(2-x^2)\beta^2}\delta H_6=0\;,\nonumber\\
		\partial_y \delta H_j=0\;,\qquad j\in\{3, 4, 6, 8, 9, 11\}\;.
	\end{gather}
Under the coordinate transformation \eqref{eq:coordtrans} there are boundary conditions at $\xi=0$ and $\xi=1$ that are equivalent to those at $y=0$ and $x=0$, respectively.
At infinity $(\rho=0)$, we impose the lowest-order series expansion of the equations about $\rho=0$.  There, the boundary conditions are too long to merit their inclusion in this manuscript.

As mentioned in the main text, we solve the equations above using Newton-Raphson.  As a means of obtaining a seed, we set the background to the static ring (\emph{i.e.} $\alpha=1$) and solve
	\begin{equation}\label{eq:negativemodefull}
		h^a{}_a=0\;,\qquad \nabla^ah_{ab}=0\;,\qquad \triangle _Lh_{ab}=-k^2h_{ab}\equiv-\frac{\widetilde k^2}{R^2{k_0}^2}h_{ab}\;.
	\end{equation}
We can take the same set of equations and boundary conditions we had before with one difference.  When $\tilde{k}\neq0$, the functions near infinity ($\rho=0$) go as $\widetilde {\delta H}\sim e^{-\widetilde k/\rho}$, so there we impose the Dirichlet conditions $\delta H_j=0$. 

Since we are on the static ring, we have $\Omega=0$.  If we further redefine $\omega=i\Gamma$, one can show that all factors of $i$ drop out of the equations of motion and boundary conditions. Hence, we are solving an eigenvalue problem with \emph{real} matrices.

To obtain our first solution, we set $\Gamma=0$ and view $\eqref{eq:negativemodefull}$ as a linear eigenvalue problem in $\tilde k^2$.  (Note that had we factored out $e^{-\widetilde k/\rho}$ from the $\delta H$'s, we would have been left instead with a quadratic eigenvalue problem in $\tilde k$.)  In this case, there are extra symmetries that allow us to set $\delta H_6=\delta H_8=\delta H_9=0$.  For a particular static ring ($\beta=1$ was a good choice), we were able to find a single positive, real value of $\widetilde k^2$ via QZ factorisation.  

Having obtained our first solution, we continue to view $\eqref{eq:negativemodefull}$ as a linear eigenvalue problem in $\tilde k^2$, but now solve it using Newton-Raphson.  We slowly increase $\Gamma$ until we find that $\tilde k$ is close to zero.  Note that if $\Gamma\neq0$, we are no longer allowed to set $\delta H_6=\delta H_8=\delta H_9=0$, and must solve for the full set of equations.

Once $\tilde k$ is close to zero, we can input the result as a seed for \eqref{eq:eomfull}, viewed as a quadratic eigenvalue problem in $\Gamma$.  We then increase $\alpha$ (and hence also $\Omega$) until the ring is balanced.  Note that when $\Omega\neq0$, we expect $\Gamma$ to now be complex.  Once balanced, we can explore the rest of the parameter space of the balanced ring.  

\begin{figure}
\begin{center}
\includegraphics[width=\textwidth]{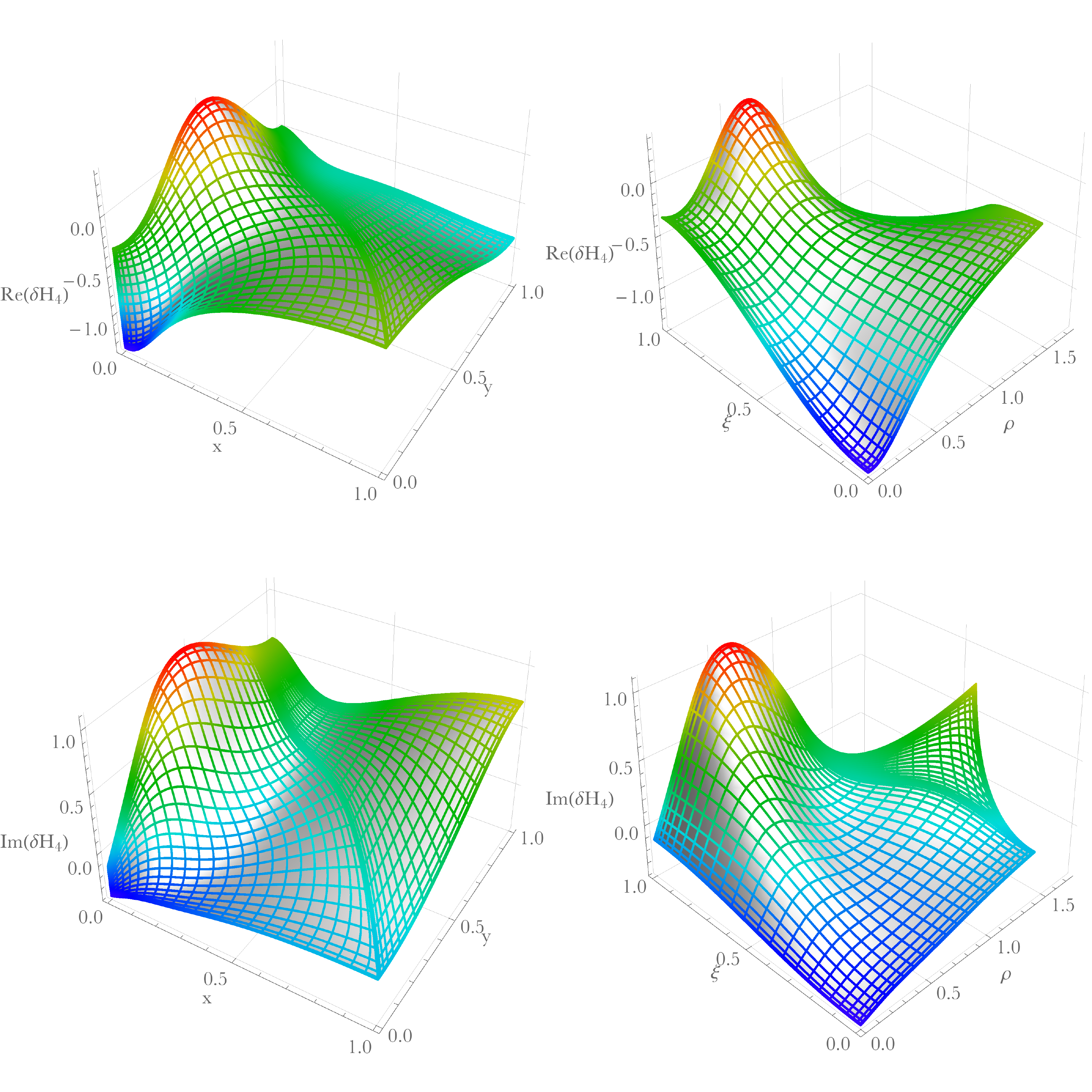}
\end{center}
\caption{Real (top plots) and Imaginary (bottom plots) parts of $\delta H_4$ in the $(x,y)$ coordinate system (left plots) and $(\rho,\xi)$ coordinate system (right plots).}\label{fig:eigenfunction}
\end{figure}

At this point, we give a few numerical results to our calculation.  A plot of the eigenfunction $\delta H_4$ is given in FIG.~\ref{fig:eigenfunction}.  The apparent smoothness of the function is reassuring.  From these plots, one can see the non-overlapping patches we have placed with transfinite interpolation.  One can also see the apparent coordinate singularity at $x=y=0$ and that it is removed when mapped to $\rho=0$ in the $(\rho,\xi)$ coordinate system.  We have chosen to normalise the eigenfunctions so that $\delta H_1=1+i$ at the corner where the horizon meets the inner axis.  

\begin{figure}
\centering
\includegraphics[width=.45\textwidth]{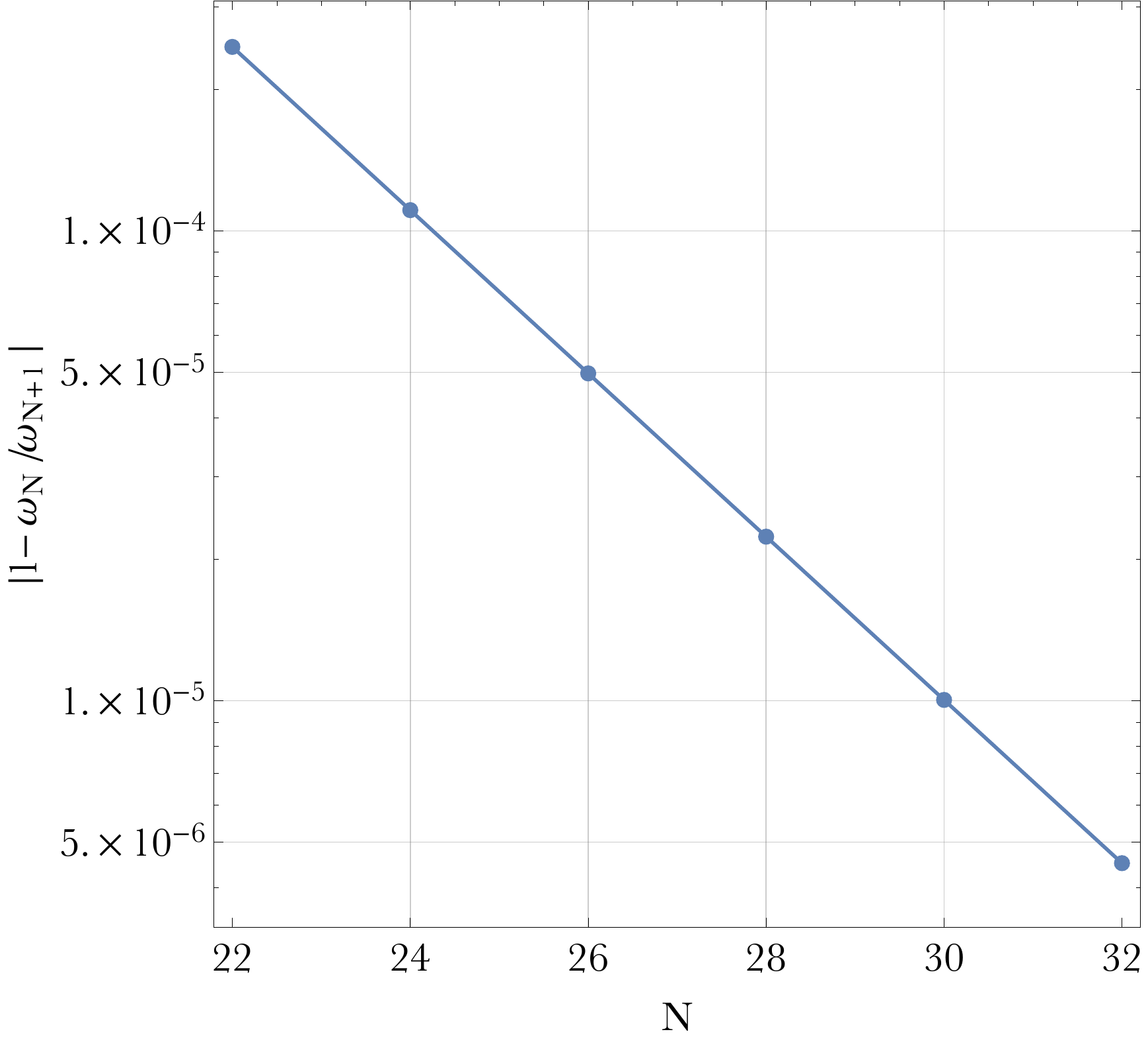}
\caption{Convergence test as a function of the number of points $N$. The resolutions used were of the form $(N+N)\times N$, indicating $N^2$ points per patch.}
\label{fig:convergence}
\end{figure}

We have performed a convergence test, the results of which are displayed in Fig.~\ref{fig:convergence}. We monitored the quantity $|1-\omega_{N}/\omega_{N+1}|$ for several different resolutions. Here, $\omega_N$ is the complex frequency of the unstable mode with $m=2$ described in the main text, computed for $\nu=1/2$. The results are plotted in a $\log$ scale in Fig.~\ref{fig:convergence}, where a straight line can be seen. This is good evidence that the convergence of our method is exponential, as dictated by spectral collocation methods.

\emph{\bf{Embedding Diagrams.}} Now let us construct an embedding diagram.  Recall that the ring has horizon topology $S^1\times S^2$.  We will embed the $S^2$ of this horizon into $\mathbb R^3$, and plot the result as a function of the $S^1$ coordinate.  This partially follows \cite{Elvang:2006dd}.

First, let us move to ingoing Eddington-Finkelstein coordinates
	\begin{equation}
		v=t+\log(1-y^2)\;,\qquad \tilde\psi=\psi+\Omega\log(1-y^2)\;.
	\end{equation}
In these coordinates, we have
	\begin{equation}
		e^{-i\omega t+im\psi}(1-y^2)^{-i(\omega-m\Omega)}=e^{-i\omega v+im\tilde\psi}\;,
	\end{equation}
so constant $v$ and $\tilde\psi$ slices on the horizon have nontrivial perturbations.  At constant $v$ and $\tilde\psi$, the induced metric on the horizon takes the form
	\begin{equation}\label{eq:horizonform}
		g_{xx}(x)dx^2+g_{\phi\phi}(x)d\phi^2\;.
	\end{equation}
Our aim is to embed the $S^2$ of the ring horizon into $\mathbb R^3$:
	\begin{equation}
		dX^2+dY^2+Y^2d\phi^2\;,
	\end{equation}
which we have chosen to write in cylindrical coordinates.  Now let us suppose we have a parametrised curve $X(x)$ and $Y(x)$.  Then the induced metric on this curve is
	\begin{equation}
		(X'(x)^2+Y'(x)^2)dx^2+Y(x)^2\phi^2\;.
	\end{equation}
Equating this with \eqref{eq:horizonform} gives
	\begin{equation}
		Y(x)=\sqrt{g_{\phi\phi}(x)}\;,\qquad X'(x)=\sqrt{g_{xx}(x)-Y'(x)^2}\;,
	\end{equation}
which is an ordinary differential equation we can solve to compute the embedding diagram.

\begin{figure}
\centering
\includegraphics[height=0.3\textheight]{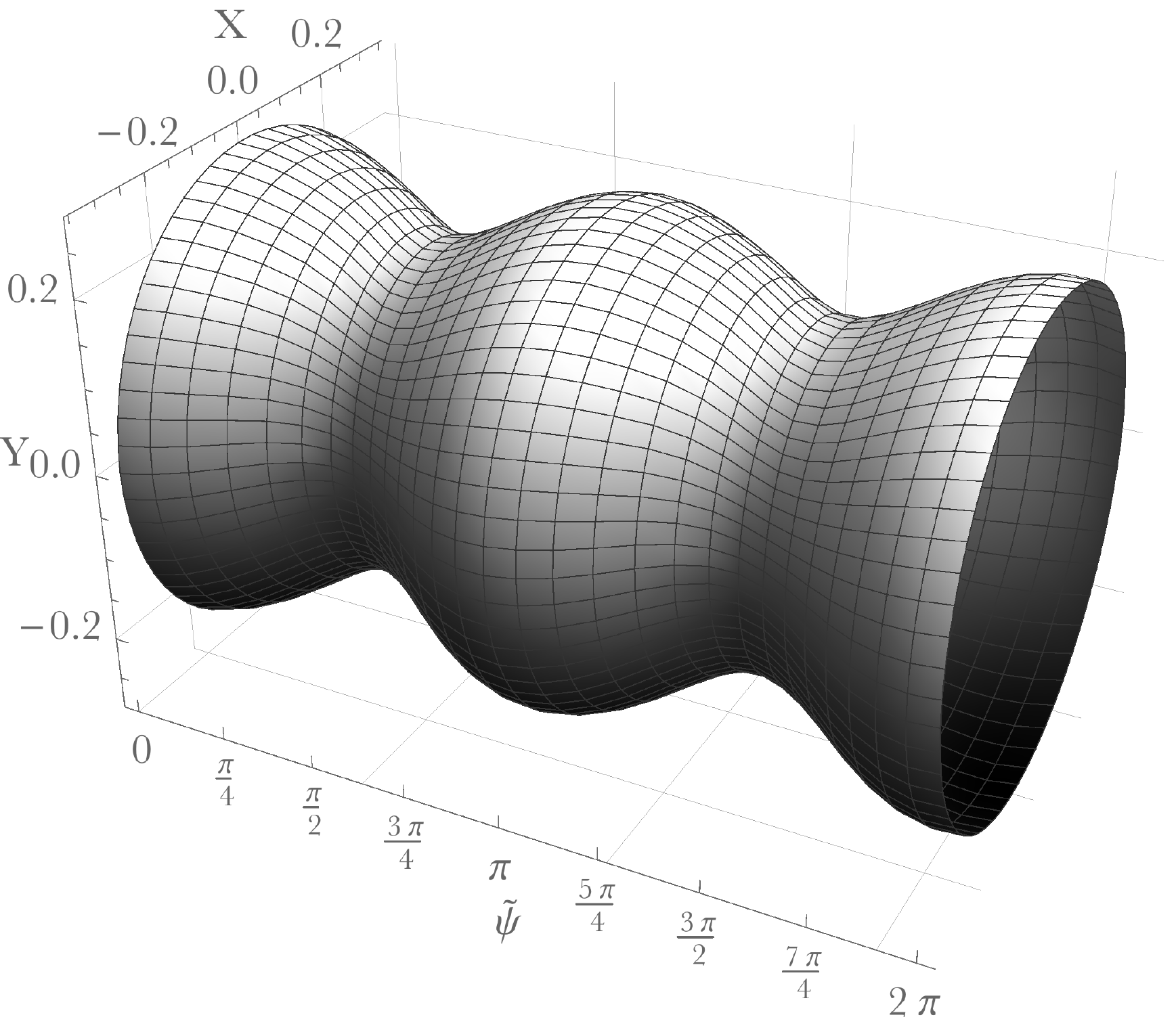}
\hspace{0.5 cm}
\includegraphics[height=0.3\textheight]{graphics/embedding2.pdf}
\caption{Isometric embedding of constant $\tilde\psi$ slices of the $S^2$ spatial sections of the perturbed black ring horizon.  The left plot corresponds to $\nu=0.5$, while the right corresponds to $\nu=0.2$.}\label{fig:embedding2}
\end{figure}

To fix units and normalisations, we rescale $GM=1$, fix the Eddington-Finkelstein coordinate $v=0$, and normalise the eigenfunctions so that $\widetilde{\delta H}_5=1$ where the horizon meets the inner axis. The integration constant above is fixed so that the $S^2$ is centred on the coordinates (\emph{i.e.} the maximum and minimum value of $X$ on the embedding differ only by a sign), with the inner axis of the horizon at negative $X$ and the outer axis at positive $X$.  Though this embedding is in $\mathbb R^3$, the symmetry about $\phi$ lets us ignore this coordinate an plot just $X(x)$ and $Y(x)$.  We do so as a function of $\tilde\psi$.  The results of this embedding are shown in FIG.~\ref{fig:embedding2} for two values of $\nu$.

\newpage\hspace{1cm}
\newpage

\smallskip
\twocolumngrid
%%%%%%%%%%%%%%%%%%%%%%%%%%%%%%%%%%%%%%%%%%%%%%%
\bibliography{refs}{}
\bibliographystyle{JHEP}

\end{document}